\documentclass[prd,12pt,aps,superscriptaddress,preprintnumbers,showpacs,tightenlines,floatfix,nofootinbib,amssymb]{revtex4}
\usepackage{epsfig}

\newcommand{\be}{\begin{equation}}
\newcommand{\ee}{\end{equation}}
\newcommand{\beqn}{\begin{eqnarray}}
\newcommand{\eeqn}{\end{eqnarray}}
\newcommand{\eq}[1]{(\ref{#1})}

\newcommand{\dd}{\mathrm{d}}

\newcommand{\dual}{\mbox{}^{\ast}}

\newcommand{\Kanazawa}{\affiliation{Institute for Theoretical Physics,
Kanazawa University, Kanazawa 920-1192, Japan}}
\newcommand{\ITEP}{\affiliation{Institute of Theoretical and
Experimental Physics, B.Cheremushkinskaya 25, Moscow, 117259, Russia}}
\newcommand{\Berlin}{\affiliation{Institut f\"ur Physik, Humboldt-Universit\"at zu Berlin,
D-12489 Berlin, Germany}}
\newcommand{\Leipzig}{\affiliation{Institut f\"ur Theoretische Physik and NTZ, Universit\"at
Leipzig, D-04109 Leipzig, Germany}}

\begin{document}

\title{The photon propagator in compact QED$_{2+1}$: \\
       the effect of wrapping Dirac strings}

\author{M.~N.~Chernodub}\Kanazawa\ITEP
\author{E.-M.~Ilgenfritz}\Berlin
\author{A.~Schiller}\Leipzig

\preprint{LU-ITP 2003/029}
\preprint{KANAZAWA/2003-31}
\preprint{ITEP-LAT-2003-27}
\preprint{HU-EP-03/78}

\begin{abstract}
We discuss the influence of closed Dirac strings on the photon propagator
in the Landau gauge
emerging from a study of the compact $U(1)$ gauge model in $2+1$ dimensions.
This gauge also minimizes the total length of the
Dirac strings. Closed Dirac strings are stable against local gauge-fixing
algorithms only due to the torus boundary conditions of the lattice.
We demonstrate
that these left-over Dirac strings are responsible for the
previously observed unphysical behavior of the propagator of space-like
photons ($D_T$) in the deconfinement (high temperature) phase.
We show how one can monitor the number $N_3$ of thermal Dirac strings
which allows to separate the propagator measurements into $N_3$ sectors.
The propagator in $N_3 \neq 0$ sectors is characterized by a non--zero
mass and an anomalous dimension similarly to the confinement phase.
Both mass squared and anomalous dimension  are found to be proportional
to $N_3$.  Consequently, in the $N_3=0$ sector the unphysical behavior of
the $D_T$ photon propagator is cured and the deviation from the free
massless propagator disappears.
\end{abstract}

\pacs{11.15.Ha, 11.10.Wx, 12.38.Gc}

\date{November 21, 2003}

\maketitle

\section{Introduction}
\label{one}

The interest in the three-dimensional compact electrodynamics (cQED$_3$)
has two roots: (i) this model shares similar features with QCD
such as confinement~\cite{Polyakov} and chiral symmetry breaking~\cite{ChSB}
and (ii) cQED$_3$ has applications to condensed matter systems such as
Josephson junction arrays~\cite{Josephson}
and high-$T_c$ superconductors~\cite{HighTc}.
All non-perturbative features of cQED$_3$ arise
thanks to the compactness of the gauge field, which, in turn, leads to the
appearance of monopoles. The monopole plasma at low temperature phase gives
rise to the confinement of electric charges~\cite{Polyakov} whereas
at high temperature the confinement disappears due to binding of monopoles
and antimonopoles into dipoles~\cite{AgasianZarembo,Chernodub:2001ws}.

The confinement property
manifests itself also in the gauge boson propagator in the Landau
gauge~\cite{Chernodub:2001mg}. The effect is twofold: first, an
''anomalous dimension'' $\alpha$ appears which modifies the momentum
dependence of the propagator, and second, a mass $m$ is generated which
can be well understood in terms of Polyakov's theory~\cite{Polyakov}.
As it is shown in Ref.~\cite{Chernodub:2001mg}, all nontrivial
effects reside exclusively in the singular fields of the monopoles.
At the critical temperature -- where the monopole plasma turns into the
dipole plasma -- both effects disappear.

The monopole binding is
observed in the zero-temperature
model, too, in the presence of matter fields.
At sufficiently strong coupling between
gauge and matter fields
the monopole plasma also turns into a dipole
plasma~\cite{AnomalousMatter,Chernodub:2002ym}. At weak coupling the
dynamical matter fields
have some influence on the anomalous dimension of
the gauge boson propagator~\cite{AnomalousMatter,Chernodub:2002en}.

Once monopoles and antimonopoles have turned into pairs they cannot contribute
to non-perturbative effects like
anomalous dimension and mass generation. Therefore it is natural to
expect that in the high temperature phase of cQED$_3$ the masses and
anomalous dimensions characterizing the photon propagator
have to vanish. However, the numerical results of Ref.~\cite{Chernodub:2002gp}
seem to indicate the existence of a non-zero mass and anomalous dimension for the propagator of the spatial photons even
in the high temperature (deconfinement) phase. It was suggested in Ref.~\cite{Chernodub:2002gp} that the spatial photons are affected by a severe Gribov copy problem which might lead to unphysical results in the Landau gauge.

The fact that the $U(1)$ gauge theory in four space-time dimensions
has a gauge fixing problem related to the Dirac
strings was discussed in Ref.~\cite{Valya:DDS}.
It was pointed out there that those gauge copies which possess
so-called double Dirac sheets (DDS) give rise to a wrong behavior of the gauge
boson propagator.

The DDS is a classical solution in the $U(1)$ gauge model~\cite{Valya:classical}
that can be considered as a world surface of a tightly bound pair of oppositely
''charged'' Dirac strings. The configuration containing a DDS
is a Gribov copy of another configuration with zero number of double sheets.
The DDS's are closed
by wrapping around the lattice torus and are not associated with any monopoles.
It was shown in Ref.~\cite{Valya:DDS} that the
practical removal of the DDS's is a quite delicate problem.

In the present paper we study the gauge boson propagator in high-temperature
cQED$_3$. We focus on gauge configurations containing Dirac strings looping along
the shortest compactified
({\it i.e.} temperature) direction. Below we shall call these loops
''thermal Dirac loops'' (TDL).
We will show that configurations without such closed Dirac loops
provide physically sane
results
(vanishing anomalous dimension and mass: $\alpha=0$, $m=0$)
for the propagator in the Landau gauge
whereas the presence of even a single closed Dirac loop leads to a non-vanishing
$\alpha$c and $m$.
We do not search for the double Dirac loops of opposite orientation (a three--dimensional
analog of the double Dirac sheet discussed in Ref.~\cite{Valya:DDS}). In our
approach those double Dirac loops
would be treated as two Dirac strings.

In Section~\ref{sec:model_etc} we recall
the lattice model and remind the tensorial structure of
the photon propagator at $T \ne 0$. In Section~\ref{sec:results} we present
the numerical results
establishing the relation between parameters of the photon propagator and
closed Dirac loops. Our conclusions are formulated in the last Section.

\section{The model and its photon propagator at finite $T$}
\label{sec:model_etc}

In this Section we briefly describe the model, the structure of the propagator
and the algorithms which were used in our work.
All these technical details are the same as described in
our earlier paper~\cite{Chernodub:2002gp}.
An interested reader may consult that paper for a more detailed description.

We use the standard Wilson single-plaquette action of cQED$_3$,
\beqn
S[\theta_l] = \beta \sum\limits_{p} \big( 1 - \cos \theta_p \big) \, ,
\label{def:action}
\eeqn
where $\theta_p$ is the $U(1)$ field strength tensor represented
by the plaquette curl of the compact link field $\theta_l$.
The lattice coupling constant $\beta$ is related to the lattice
spacing $a$ and the continuum coupling constant $g_3$
-- which has dimension ${\mathrm{mass}}^{1 \slash 2}$  --
of the $3D$ theory as follows:
\beqn
\beta = 1 \slash (a\, g^2_3)\; .
\label{def:beta}
\eeqn

The lattice corresponding to finite temperature is asymmetric, $L^2_s\times
L_t$, $L_t \ll L_s$. In the limit $L_s \to \infty$, the temporal extension
of the lattice is related to the physical temperature, $L_t = 1 \slash (T a)$.
Using (\ref{def:beta}) the temperature can be expressed in units of $g_3^2$
in the following way:
\beqn
\frac{T}{g^2_3} = \frac{\beta}{L_t} \, .
\label{def:temperature}
\eeqn
Thus the temperature $T$ is proportional to the lattice coupling $\beta$:
the low-temperature (confinement) phase is realized at small values of
$\beta$, the high-temperature (deconfinement) phase corresponds to
large $\beta$.

All our simulations are performed on a $32^2 \times 8$ lattice. For this lattice
the phase transition happens at~\cite{Chernodub:2001ws} $\beta_c=2.30(2)$.
In the confinement phase the density of the monopoles is relatively high.
The gauge dependent Dirac string bits through Dirac plaquettes defined below
(and needed to construct the gauge independent monopoles)
form either connected clusters of open Dirac strings with monopoles and
antimonopoles at their ends or clusters of closed strings.
Therefore, for a high monopole density the number of Dirac strings is large, too.

In this paper we are interested in the quantitative effects of the temporal
Dirac strings. Therefore we would prefer to work at low monopole densities
in order to be able to easily separate closed and open Dirac strings
unambiguously.
We are going to present a quantitative demonstration that thermal Dirac loops
create an unphysical behavior of the transverse photon propagator
due to our attempt to fix the configuration to Landau gauge.
Thus, we have chosen to illustrate this by a simulation at $\beta=2.6$ which
is located already deep in the deconfinement phase. At this value of $\beta$
the density of the monopoles is quite low,
$\rho_{\mathrm{mon}} = 1.5(2)\cdot 10^{-4}$,
compared to the confinement phase (where, for example,
$\rho_{\mathrm{mon}} = 0.1950(1)$ at $\beta=1.0$).
We have mentioned already that the construction of the Dirac string is gauge dependent.
An ideal Landau gauge fixing makes the open strings between the remaining
monopoles and antimonopoles (in the form of dipoles) straightened, whereas
the closed ones, if not ``wrapping'', will collapse.

The final discussion of the photon propagator will be given in lattice
momentum space.
With the vector potential $A_{{\vec x},\mu}$ defined in a specified gauge,
the propagator is written in
terms of the Fourier transformed gauge potential,
\beqn
  \tilde{A}_{{\vec k},\mu} = \frac{1}{\sqrt{L_1~L_2~L_3}}
  \sum\limits_{{\vec n}}
  \exp \Bigl( 2 \pi i~\sum_{\nu=1}^{3} \frac{k_{\nu}
  (~n_{\nu}+\frac{1}{2}\delta_{\nu\mu}~) } {L_{\nu}} \Bigr)
  ~A_{{\vec n}+\frac{1}{2}{\vec \mu},\mu} \; ,
  \label{def:fourier_transformation}
\eeqn
which is a sum over a certain discrete set of points
${\vec x}={\vec n}+\frac{1}{2}{\vec \mu}$ forming the support of
$A_{{\vec x},\mu}$ on the lattice.
These are the midpoints of the links in $\mu$ direction.
Here ${\vec n}$ denotes the lattice sites (nodes) with integer Cartesian
coordinates. The propagator is the gauge-fixed ensemble average
of the following bilinear in $\tilde{A}$,
\beqn
  D_{\mu\nu}({\vec p}) = \langle \tilde{A}_{ {\vec k},\mu}
                               \tilde{A}_{-{\vec k},\nu} \rangle \; .
  \label{def:propagator}
\eeqn
We use the {\it sine}-definition of the gauge potential:
\beqn
  A_{{\vec n}+\frac{1}{2}{\vec \mu},\mu}
    = \sin \left( \theta_{{\vec n},\mu} \right)/(g_3~a)
    = \left( U_{{\vec n},\mu} - U^{*}_{{\vec n},\mu} \right)/(2~i~g_3~a) \, .
  \label{def:sine}
\eeqn
The lattice momenta ${\vec p}$ on the left hand side of (\ref{def:propagator})
are related to the integer valued Fourier momenta ${\vec k}$ as follows:
\beqn
  p_\mu(k_\mu)=  \frac{2}{a} \sin \frac{ \pi k_\mu}{L_\mu}
  \,, \quad  k_\mu=0, \pm 1,..., \pm \frac{L_\mu}{2} \; .
  \label{def:momenta}
\eeqn

In the finite temperature case, the propagator can be
parameterized by three scalar functions (formfactors),
\beqn
  D_{\mu\nu}({\vec p})= P^T_{\mu\nu}({\vec p}) D_T(|{\mathbf p}|,p_3)
                   + P^L_{\mu\nu}({\vec p}) D_L(|{\mathbf p}|,p_3)
              + \frac{p_\mu p_\nu}{p^2} \frac{F(|{\mathbf p}|,p_3)}{p^2}\,,
  \label{def:all_components_at_T}
\eeqn
where we have identified the temperature direction with the $\mu=3$ axis, and
have introduced the notations ${\mathbf p}^2 = p_1^2 + p_{2}^2$ and
$|{\mathbf p}|=\sqrt{{\mathbf p}^2}$. If the Landau gauge is exactly fulfilled,
one would expect that  $F(p^2) \equiv 0$.
In our simulations presented below this function is
indeed very close to zero.

Eq.~\eq{def:all_components_at_T} contains also two projection operators,
namely, the transverse (with respect to the temperature direction)
projection operator $P^T$ and the two-dimensional longitudinal projection
operator $P^L$, respectively (with $i,j=1,2$),
\beqn
  P^T_{ij}({\vec p}) & = & \delta_{ij}- \frac{p_i p_j}{{\mathbf p}^2} \,, \quad
  P^T_{33}({\vec p}) = P^T_{3i}({\vec p}) = P^T_{i3}({\vec p}) = 0 \, ,
  \label{def:PT}
  \\
  P^L_{\mu\nu}({\vec p}) & = & P_{\mu\nu}({\vec p}) - P^T_{\mu\nu}({\vec p})\,,    \quad
  P_{\mu\nu}({\vec p}) = \delta_{\mu\nu}- \frac{p_\mu p_\nu}{p^2}\,.
  \label{def:PL}
\eeqn

In the static limit, $p_3=0$, the scalar function $D_L$ is equivalent
to the correlator of temporal photons $D_L(|{\mathbf p}|,p_3=0)
\equiv D_{33}(|{\mathbf p}|,p_3=0)$. The properties of this type of propagator
have been discussed in Ref.~\cite{Chernodub:2001mg}. Analogously, the scalar
function $D_T$ describes the behavior of the spatial photons.
The behavior of {\it this} formfactor $D_T$ is of special interest in
the present paper.

In order to analyze the effect of the Dirac strings on the propagator in
general we separate
singular (monopole) and regular (photon) contributions to the lattice gauge field
on the level of the link angles $\theta_l$ following
Refs.~\cite{PhMon,Chernodub:2001mg,Chernodub:2002gp,Chernodub:2002en},
\beqn
  \theta = \theta^{\mathrm{phot}} + \theta^{\mathrm{mono}}\,, \quad
  \theta^{\mathrm{mono}} = 2 \pi \Delta^{-1} \delta p[j]\,,
  \label{def:form_theta_decomposition_1}
\eeqn
where the dual zero-form $\dual j$ represents the monopoles on the
dual lattice sites ($i.e.$ the monopoles are defined on the cubes
of the original lattice), $\Delta^{-1}$ is the inverse lattice Laplacian.
The one-form on the dual lattice $p[j]$ defines the
Dirac strings that connect monopoles and antimonopoles
because of the condition $\delta \dual p[j] = \dual j$.

The photon part $\theta^{\mathrm{phot}}$ is free of singularities whereas the monopole part $\theta^{\mathrm{mono}}$ contains the information about
the monopole and Dirac string singularities:
\beqn
 \frac{1}{2\pi} \dd {[\dd \theta^{\mathrm{phot}}]}_{2\pi} = 0 \,, \quad
 \frac{1}{2\pi} \dd {[\dd \theta^{\mathrm{mono}}]}_{2\pi} = j \, .
 \label{def:form_theta_decomposition_2}
\eeqn
Here the DeGrand-Toussaint definition of the monopole~\cite{DGT} was used.
Thus, besides the total propagator~\eq{def:propagator} below
we will also study the singular contribution to the propagator,
$D^{\mathrm{mono}}({\vec p}) = \langle \tilde{A}^{\mathrm{mono}}_{ {\vec k},\mu}
\tilde{A}^{\mathrm{mono}}_{-{\vec k},\nu} \rangle$, and
regular contribution $D^{phot}_{\mu\nu}({\vec p})
= \langle \tilde{A}^{phot}_{ {\vec k},\mu} \tilde{A}^{phot}_{-{\vec k},\nu} \rangle$.
The total contains also the mixed contribution $D^{mixed}_{\mu\nu}({\vec p}) =
\langle \tilde{A}^{phot}_{ {\vec k},\mu} \tilde{A}^{mono}_{-{\vec k},\nu} \rangle$,
which is not explicitly studied in this paper.

It turns out that the momentum dependence of the formfactors
$D_L$ and $D_T$
forming the total propagator can accurately be described
in both phases by the
functional form~\cite{Chernodub:2001mg,Chernodub:2002gp,Chernodub:2002en},
\beqn
  D(p^2) = \frac{Z}{\beta}\frac{m^{2\alpha}}{p^{2(1+\alpha)}+m^{2(1+\alpha)}}
  + C\,,
  \label{def:anomalous_fit}
\eeqn
where $\alpha$ is an anomalous dimension and $m$ a mass parameter;
$Z$ represents the renormalization of the photon wave-function and $C$ corresponds
to a point-like interaction between the photons which we relate to a lattice
artifact. In the following we denote the fit parameter for the $D_{L/T}$ formfactors by $\alpha_{L/T}$ etc.

\begin{figure}[!htb]
  \begin{center}
    \begin{tabular}{cc}
    \includegraphics[angle=0,scale=0.35,clip=true]{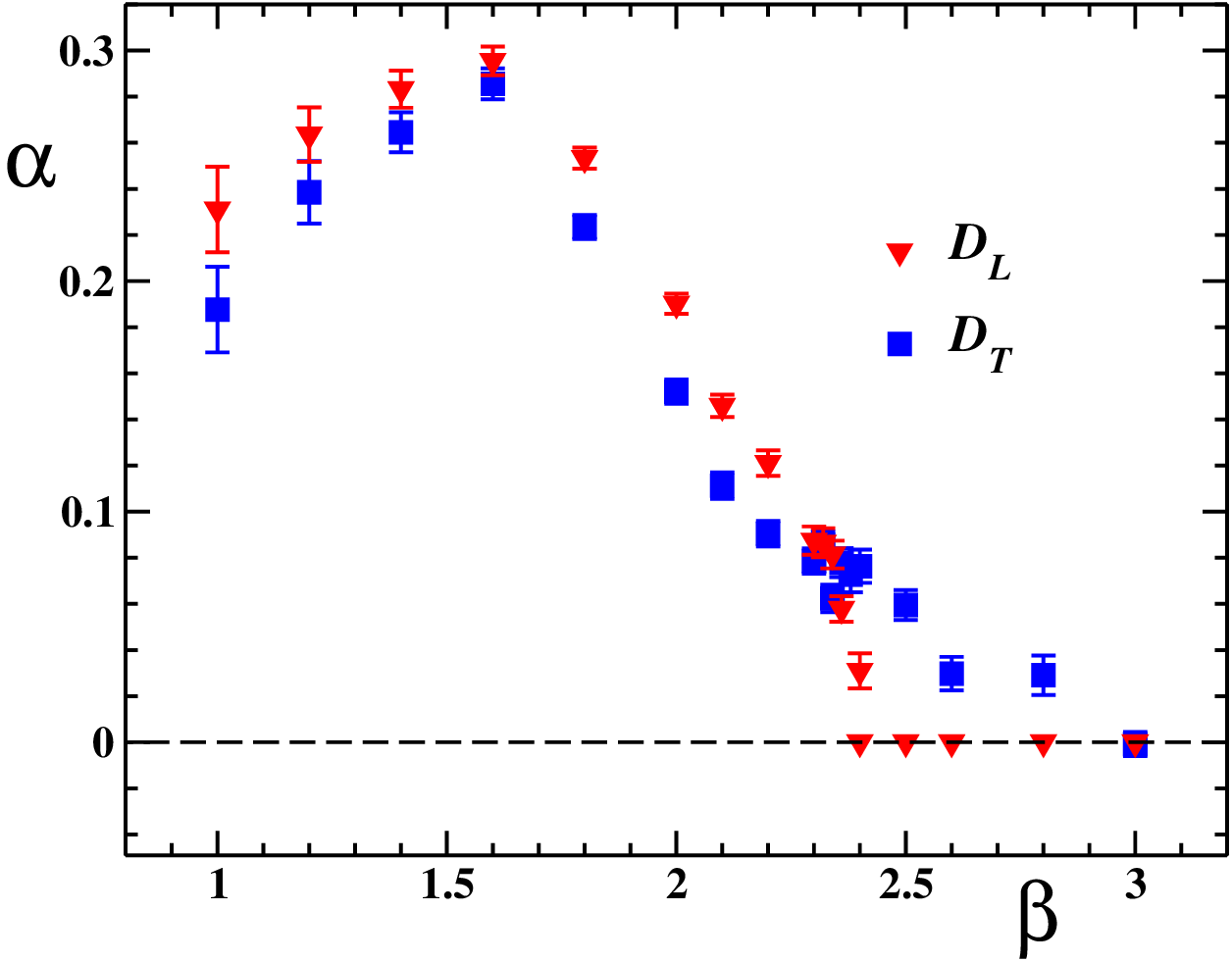} &
      \hspace{5mm}
    \includegraphics[angle=0,scale=0.35,clip=true]{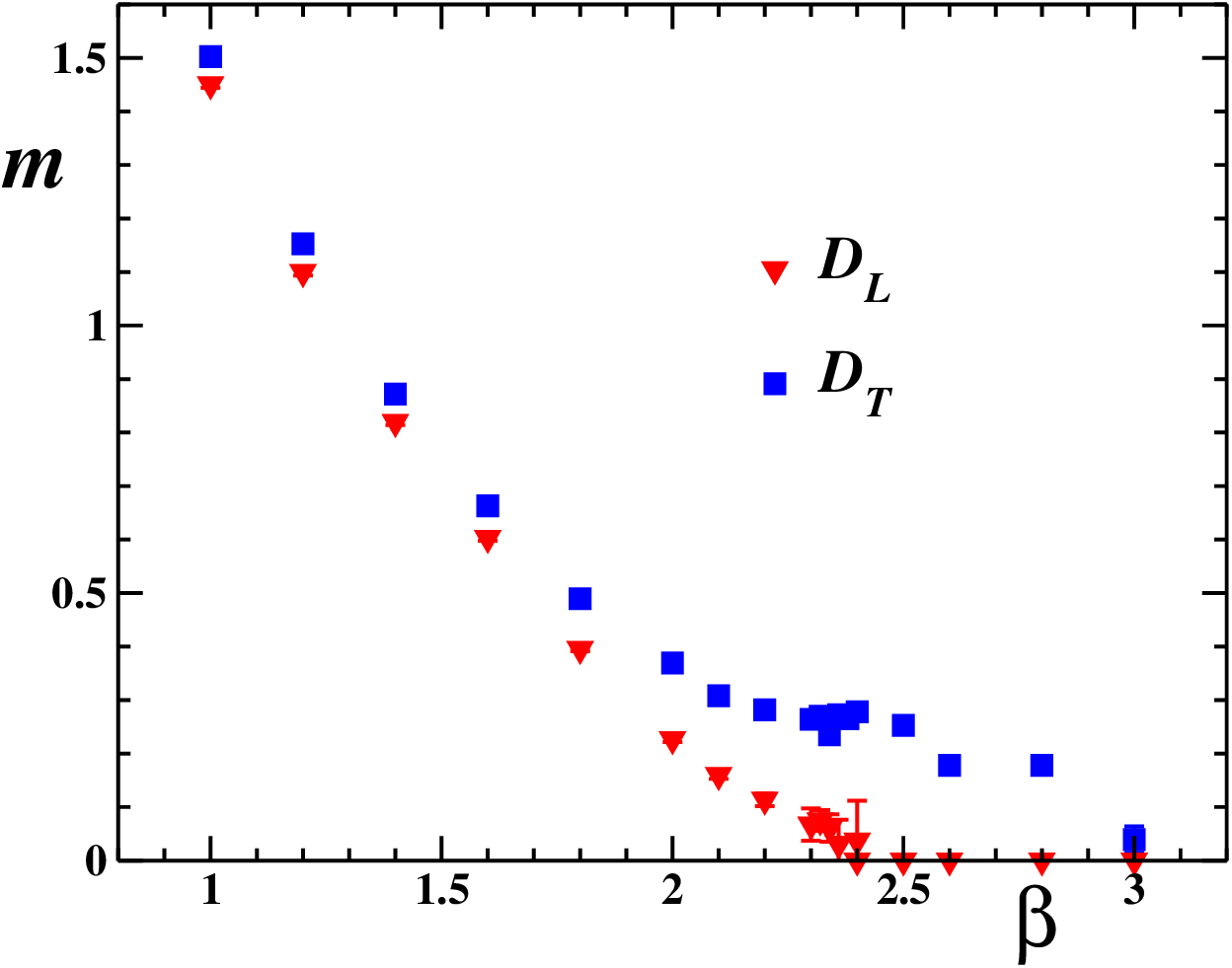} \\
      (a) & \hspace{5mm}  (b)
    \end{tabular}
  \end{center}
 \caption{Fit parameters $\alpha_{L/T}$ and $m_{L/T}$ vs.
  $\beta$ for $D_L$ and $D_T$ formfactors
  (from Refs.~\cite{Chernodub:2001mg,Chernodub:2002gp}).}
  \label{fig:old}
\end{figure}

In Figure~\ref{fig:old} we present the fitting results for the $D_L$ and $D_T$
formfactors, which were obtained in Refs.~\cite{Chernodub:2001mg,Chernodub:2002gp},
as functions of $\beta$ (temperature). The formfactors
were studied in the static limit, $p_3=0$, and then fitted by the
function~\eq{def:anomalous_fit} where $p^2$ was identified with ${\mathbf p}^2$.
%
%
One can see from Figure~\ref{fig:old}(a) that the anomalous dimension for the
temporal photons, $\alpha_L$, vanishes exactly at the point of the phase transition
($\beta_c=2.30(2)$ according to Ref.~\cite{Chernodub:2001ws})
whereas the spatial photons do not seem to feel this transition.
This is documented by $\alpha_T$ not vanishing at $\beta=\beta_c$.
Similarly, a different behavior
is seen for the mass parameters $m_L$ and $m_T$ as shown in Figure~\ref{fig:old}(b).

The vanishing of the parameters $\alpha_L$ and $m_L$ of $D_L$ at
$\beta_c$ and beyond is clearly
corroborating the finite temperature
phase transition~\cite{Chernodub:2001mg} caused by dipole formation.
The fields of the dipoles are weak at large distances and they cannot
cause neither the Debye screening~\cite{no-screening} nor
confinement\footnote{The dipole gas modifies only a short-distance interaction
between electric charges providing a small linear correction to the
Coulomb interaction~\cite{DipoleGas}.}. Thus the origin of the non-zero mass $m_T$ of the spatial photon propagator in the deconfinement phase
is likely to be an artifact of the lattice simulations,
more specifically, of the insufficient gauge fixing.
We have checked in Ref.~\cite{Chernodub:2002gp} that this propagator is
strongly affected by the Gribov copy problem.
As we have argued in Ref.~\cite{Chernodub:2002gp},
the Landau gauge minimizes the total length of the Dirac
strings and, in the idealized case, does not allow for closed Dirac strings
(Dirac loops) to exist.
However, if the Dirac loop is closed by wrapping around the torus, it is practically impossible to get rid of this gauge artifact
relying only on local gauge-fixing algorithms. Gauge copies with unremoved
TDL's correspond to local minima of the gauge fixing functional.
We have qualitatively noticed in Ref.~\cite{Gargnano}
that these strings should be blamed for the
unphysical behavior of $D_T$ in the deconfined phase if compared to $D_L$.

\section{Quantifying the effect of wrapping Dirac strings}
\label{sec:results}

Let us now discuss the effect of closed Dirac loops on the spatial
photon propagator in a more quantitative way.
As we have already mentioned, we
have simulated the model at $\beta=2.6$ using a $32^2\times 8$ lattice.
As in our previous work we have considered
two possible update algorithms: one purely local one
(five-hit Metropolis update alternating with a microcanonical sweep)
and another which included random offers
of changing the flux in one of the three directions by one unit, augmented
by a Metropolis acceptance check. Finally, in order to reduce the influence
of choosing between different Landau gauge-fixed copies we have performed
the gauge fixing procedure 100 times on random gauge copies of the same
Monte Carlo configuration ($N_{\rm copy}=100$).
The evaluation of the propagator was done on the ''best'' configuration
corresponding to the relative maximum of the gauge functional among
all $101$ gauge fixed configurations.

In our previous work~\cite{Chernodub:2002gp}
the fit parameters which should describe the formfactor $D_T$
have been obtained for a number $N_{\rm copy}$ which was gradually increased,
and even 100 Gribov copies were found to be insufficient for convergence.
This seems to exclude the possibility to improve the result further by local
gauge-fixing algorithms exclusively.
The parameters for the longitudinal formfactor $D_L$ were found to converge
already for $N_{\rm copy}\approx 10$.
This explains why, for the present purpose, we kept $N_{\rm copy}=100$.
Due to the asymmetry of the lattice it is relatively easy that Dirac strings
are generated running around the lattice in temporal direction.
We restrict ourselves to a string search in that
direction in order to separate our ensemble of (locally) gauge-fixed configurations into classes according to the number of TDL's.
A Dirac loop is formed by a connected sequence of dual links carrying
directed Dirac string bits $n_{{\vec x};\mu}^{\rm Dirac} \ne 0$.
A full Dirac string is defined by penetrating a stack of Dirac plaquettes.
A plaquette, say $P = P_{{\vec x};\mu\nu} = P_{(x_1,x_2,x_3);1,2}$, is
identified as one of the pierced, {\it i.e. Dirac} plaquettes if
\be
   n_{{\vec x};3}^{\rm Dirac}
   = \frac{1}{2 \pi} [\theta_P]_{{\rm mod} 2 \pi}
   \,.
\ee
is a non-vanishing integer.
Positive or negative values -- usually  plus or minus one -- define the
direction of the string bit.

The implementation of such a search, following the stack of Dirac plaquettes
and counting the number of closed Dirac loops, is too time consuming in general.
Deeper in the deconfined phase, it is possible to
characterize a given thermalized and gauge fixed
configuration by summing up the modules $N_3$
and the values of those integer Dirac plaquettes $I_3$
pointing into
the third (''short'') compactified direction
\be
     N_3 ={\rm Integer} \left[\frac{1}{L_3} \sum_x
         |n_{{\vec x},3}^{\rm Dirac}| \right]
\ee
and
\be
     I_3 =
     {\rm Integer} \left[\frac{1}{L_3} \sum_x
         n_{{\vec x},3}^{\rm Dirac}\right]
          \,.
\ee
The different sectors of possible thermal Dirac loops are
labelled by a number $N_3=0, 1, \dots $
If the expected TDL's are mainly static
({\it i.e.} already minimized in length), the quantity $N_3$
counts the number of those strings under the assumption that
the number of monopoles is already very low (what is the case at $\beta=2.6$).
In case $|I_3|$ coincides with $N_3 > 1$, all TDL's have the same wrapping
orientation, otherwise TDL's with different orientation are present.
With this procedure we cannot assess whether those strings are really
''at rest'' on the same $(x_1,x_2)$  position or not, but this classification
is robust enough to allow for some local dislocations of the TDL's.
The ratio
\beqn
  r_3=\frac{ {\rm {Number \, of \, measurements \, with\,}} |I_3|=N_3}
       { {\rm {All \, measurements \, in \, sector\, }} N_3}
  \label{eq:r3}
\eeqn
counts the fraction of such strings having the same wrapping orientation.
Having identified the ''thermal Dirac loop number'' $N_3$, we can classify
a given gauge-fixed configuration according to its string content
and measure the photon propagator separately in the different sectors.

In our previous Monte Carlo analysis~\cite{Chernodub:2002gp} we have used
both local and ``blended'' updates, the latter including
in addition to the local update adding and subtraction of fluxes,
among them fluxes through the $(x_1,x_2)$-plane.
Therefore, in the light of the string sector classification, we also have
to consider the effect of the chosen Monte Carlo algorithm on this
classification.

\begin{figure}[!htb]
  \begin{center}
    \begin{tabular}{cc}
    \includegraphics[angle=0,scale=0.35,clip=true]{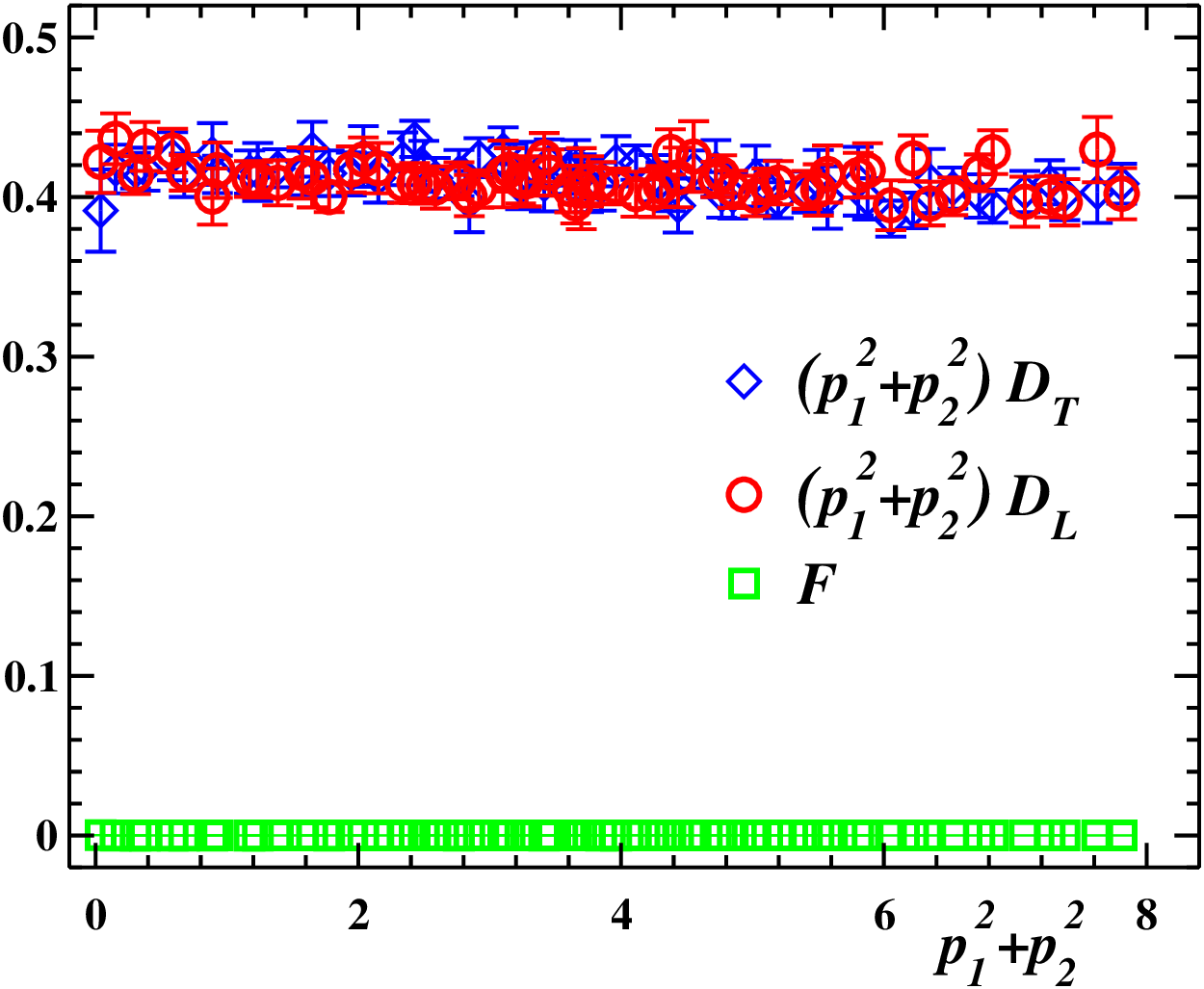} &
      \hspace{5mm}
    \includegraphics[angle=0,scale=0.35,clip=true]{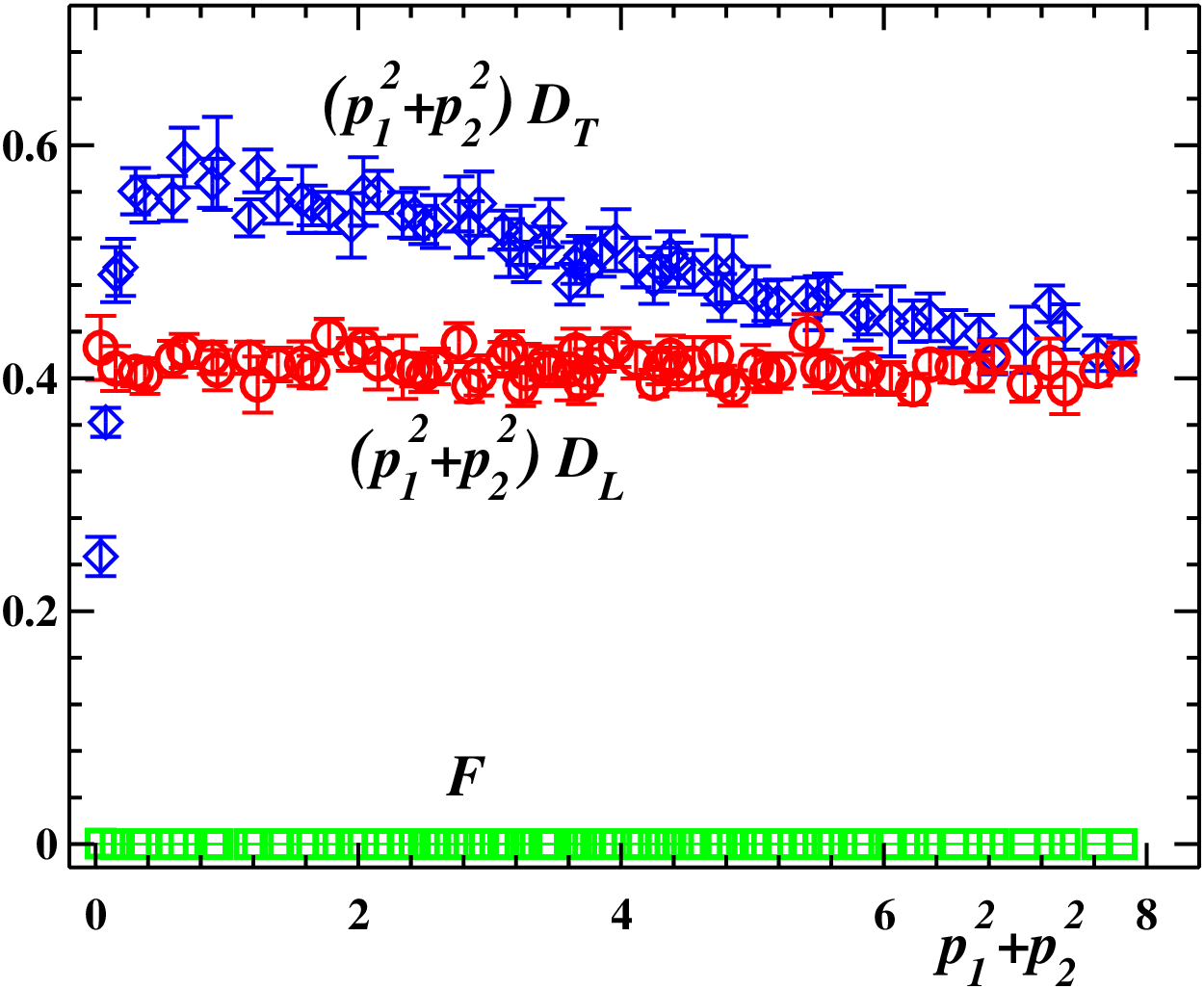}
    \\
      (a) & \hspace{5mm}  (b) \vspace{3mm}\\
    \includegraphics[angle=0,scale=0.35,clip=true]{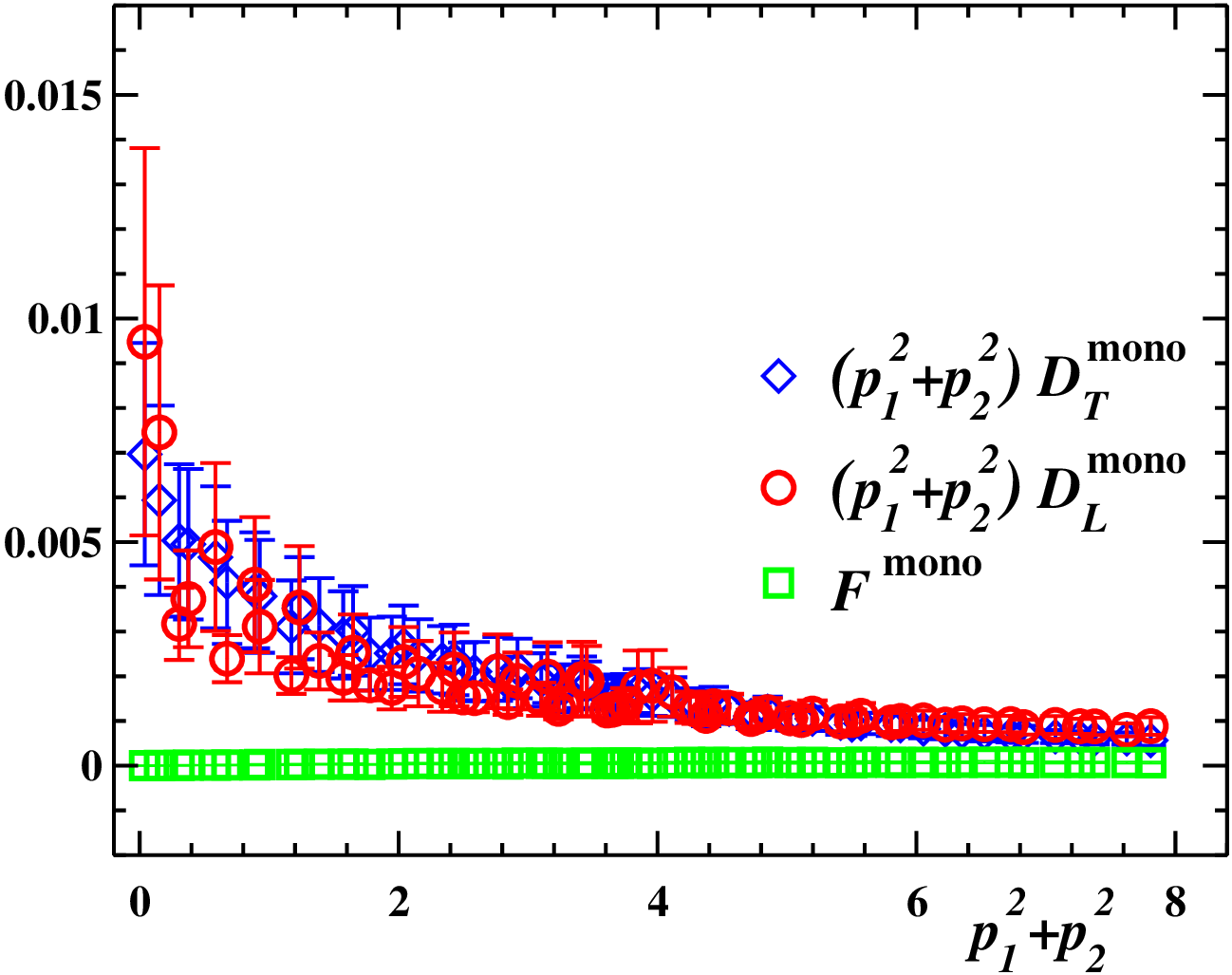} &
      \hspace{5mm}
    \includegraphics[angle=0,scale=0.35,clip=true]{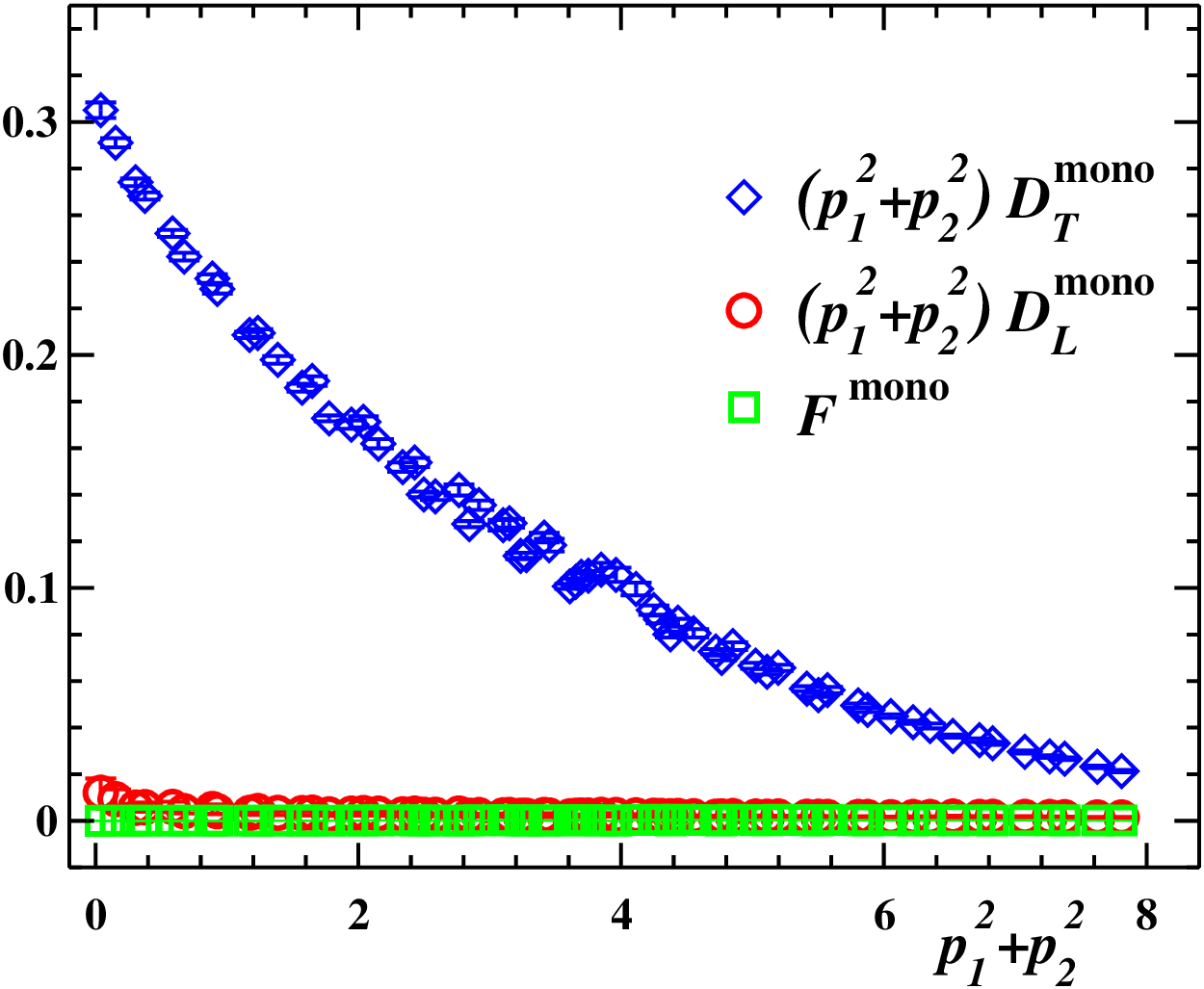}
     \\
    \includegraphics[angle=0,scale=0.35,clip=true]{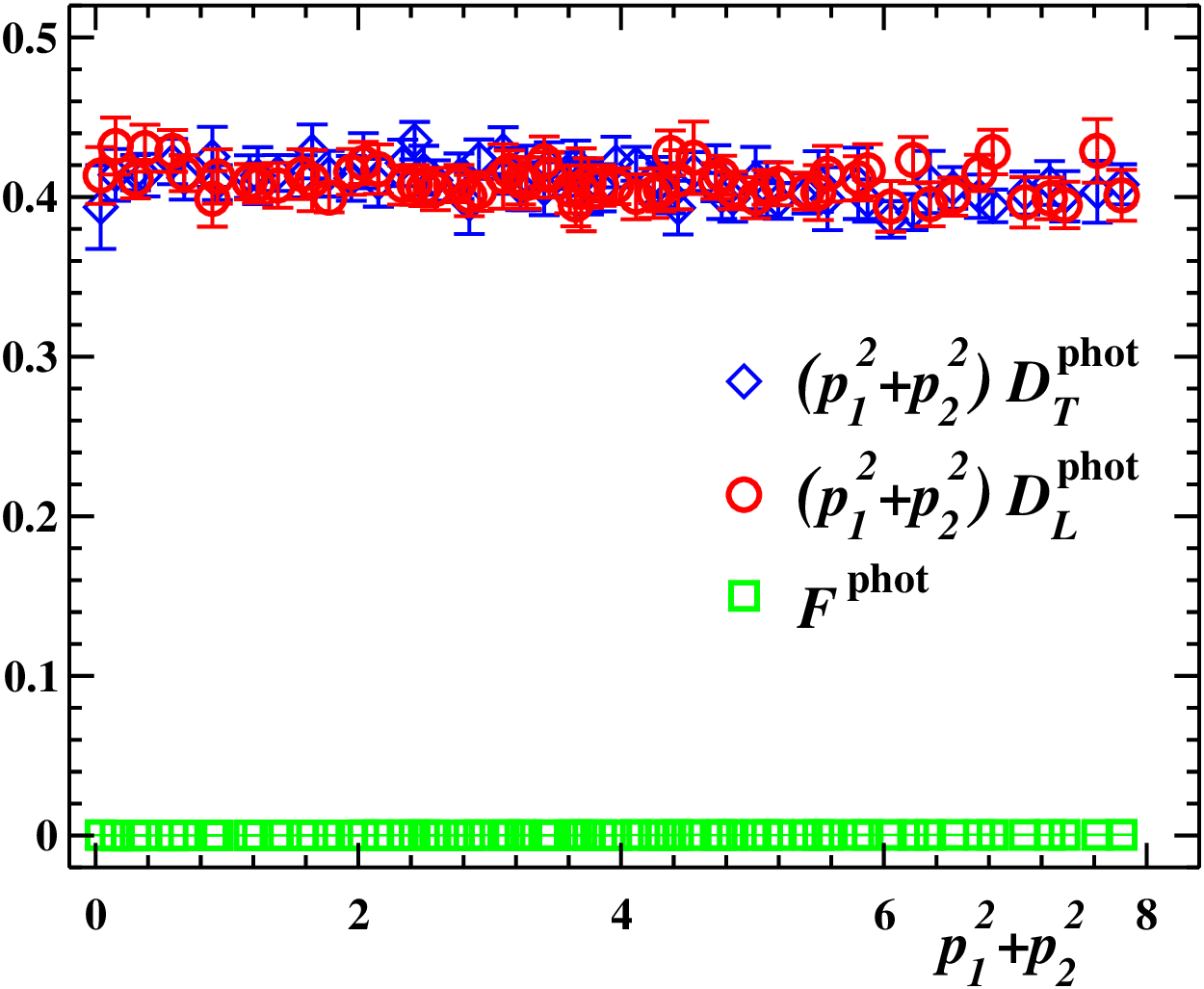} &
      \hspace{5mm}
    \includegraphics[angle=0,scale=0.35,clip=true]{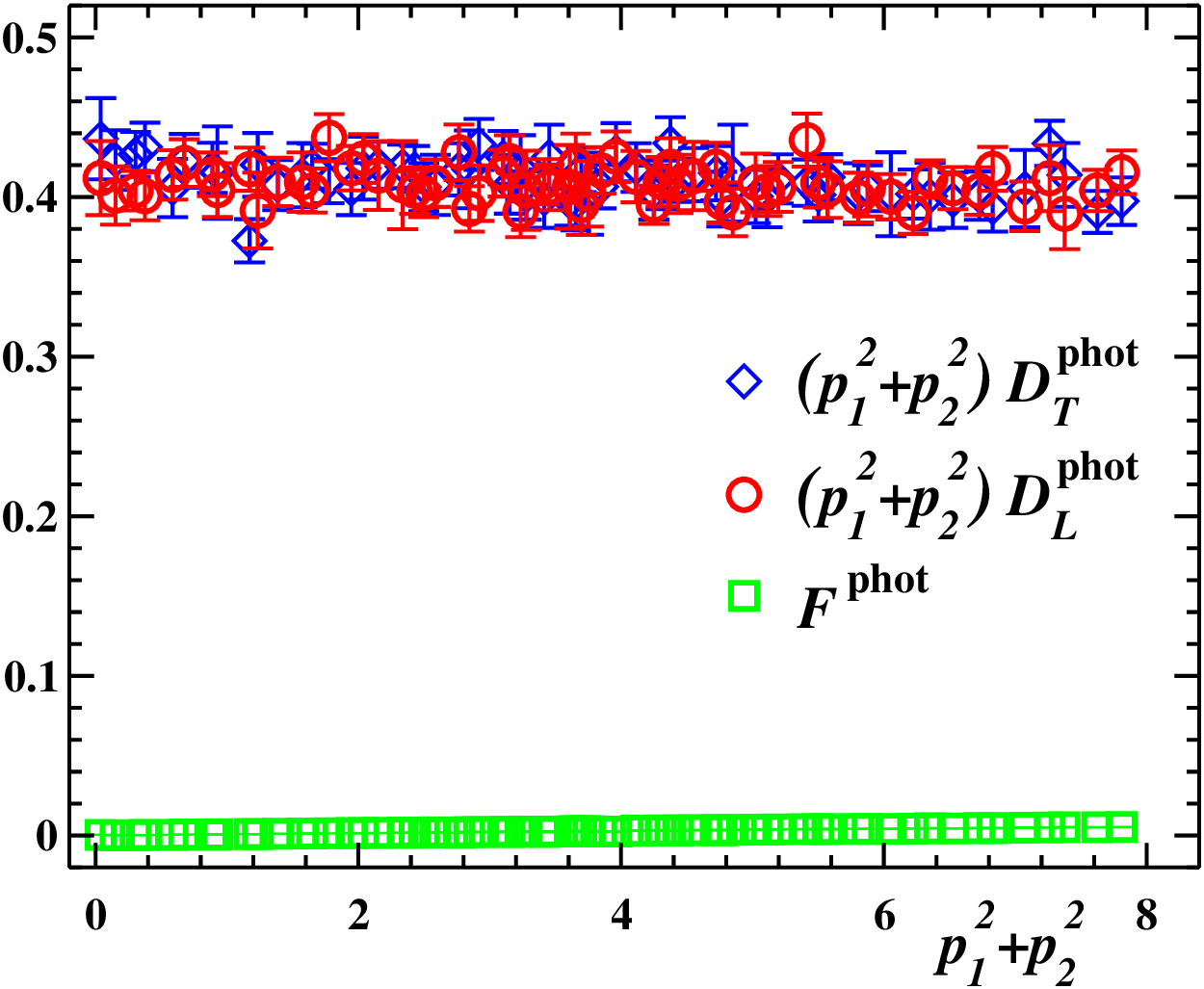}
    \\
      (e) & \hspace{5mm}  (f)
    \end{tabular}
  \end{center}
  \caption{The formfactors $D_L$, $D_T$ and $F$ {\it vs.} spatial momentum
  in the sector with (left column) zero, $N_3=0$, and (right column) one,
  $N_3=1$, thermal Dirac loop using the local update.
  The total (a,b), singular (c,d) and regular (e,f) contributions are shown.}
  \label{fig:propagator}
\end{figure}

First, we have checked the dependence of the propagators on the number
of thermal Dirac loops. In Figure~\ref{fig:propagator}
%
%
we present the formfactors $D_L$, $D_T$ and $F$ in the sectors
without and with
one TDL. Here we have used only the local update algorithm.
One may readily notice that the total formfactor of the spatial photons
$D_T$ shown in Figure~\ref{fig:propagator}(a,b)
depends significantly on the number of thermal Dirac loops
whereas the $D_L$ and $F$ formfactors are insensitive (within error bars) to
that number.

A very similar effect is observed for the singular contributions to
formfactors which are depicted in Figures~\ref{fig:propagator}(c,d).
The singular contribution
to the $D_T$ formfactor in the
sector with one TDL (Figure~\ref{fig:propagator}(d))
is about two orders of magnitude
larger than in the $N_3=0$ sector,
shown in
Figure~\ref{fig:propagator}(c). The singular contributions to other formfactors
are not affected by the presence of TDL's.

In order to show that the thermal Dirac loops make contributions only to the
singular and mixed part of the $D_T$ propagator,
whereas the regular part is not affected by the Dirac string, we plot in
Figures~\ref{fig:propagator}(e,f) the regular part of the $D_T$, $D_L$ and $F$
formfactors as a function of spatial momentum. One can see that the regular
contributions in the zero-loop and one-loop sectors
coincide within error bars.

Qualitatively, the results in Figures~\ref{fig:propagator} can be understood as
follows. The fact that the regular contribution to the propagator is insensitive to the number of TDL's is very natural since the regular part does not receive
contributions from singular structures like the Dirac strings.
The sensitivity of the $D_T$ formfactor to the number of TDL's and the respective insensitivity of the $D_L$ formfactor follows
from Eq.~\eq{def:form_theta_decomposition_1}. Indeed,
the TDL is described by a chain of the plaquettes which
are perpendicular to the temporal direction. The boundary of the
plaquette $\delta p[j]$ is a vector field the spatial components of which are
non-zero, whereas the temporal ones are vanishing. The inverse Laplacian in
Eq.~\eq{def:form_theta_decomposition_1} does not mix these components.
Thus, a thermal Dirac loop may provide a contribution only to the spatial components of the gauge potential, {\it i.e.} to $D_T$ only. Finally, the
insensitivity of the formfactor $F$ to the presence of TDL's
follows from the fact that $F$ corresponds to the longitudinal
(in momentum) part of the propagator whereas the contribution from the TDL is transverse (see Eq.~\eq{def:form_theta_decomposition_1}).

In the sector without thermal Dirac loops ($N_3=0$) the tiny singular contribution to the propagators $D_T$ and $D_L$ can be explained as an effect of remaining monopole-antimonopole pairs which are
mainly oriented in temporal direction (and/or additionally of strings in
spatial direction on which we did not trigger).

So far, the results in Figures~\ref{fig:propagator} were obtained from the
evaluation of gauge-fixed configurations when the original ensemble was
generated by exclusively local updates.
We have repeated the same analysis for the blended update algorithm mentioned
which includes also global changes of fluxes.  On this basis, we notice
that (within error bars) the results for the propagator formfactors in
the individual TDL sectors are independent of whether
blended updates are allowed or not.

However, the choice of the update algorithm (blended or local)
influences the relative weight of the individual sectors within the gauge-fixed
ensemble. Strictly speaking, the content of wrapping Dirac strings is a gauge
artifact. We are only monitoring it. Only dedicated, ''big'' gauge
transformations would be able to {\it remove} them. For example, if
global flux changes are accepted in the blended update then after applying
local gauge fixing the sector with $N_3=\pm 1$ is slightly dominating.
If a local update is applied without offering global flux changes,
after local gauge fixing the $N_3=0$ sector is clearly the dominating one.
An overview of the statistics available for this study is given in
Table~\ref{tbl:statistics}.
\begin{table}[!htb]
\begin{center}
\vspace{0.5cm}
\begin{tabular}{||l|c|c|c||}
\hline
$N_3 $ (Update) & $r_3$ & $\#$ of meas. in sector $N_3$&$\#$ of meas. attempts \\
\hline
   0 (local)   &   1.0   &      1191  &  2000  \\[0.7ex]
   0 (blended) &   1.0   &       767  &  2000  \\[0.7ex]
   1 (local)   &   0.891 &       911  &  2500  \\[0.7ex]
   1 (blended) &   0.903 &      1115  &  2500  \\[0.7ex]
   2 (local)   &   0.866 &      1912  &  7000  \\[0.7ex]
   2 (blended) &   0.778 &       974  &  7000  \\[0.7ex]
   3 (local)   &   0.859 &       580  & 15000  \\[0.7ex]
   3 (blended) &   0.869 &       314  & 20500  \\[0.7ex]
   4 (local)   &   0.895 &        19  &600000  \\[0.7ex]
   4 (blended) &   0.931 &       72   &600000  \\
\hline
\end{tabular}
\end{center}
\caption{Statistics for the  different thermal Dirac loop sectors
   using local or blended update.}
\label{tbl:statistics}
\end{table}

{}From this Table we can speculate that in the process of gauge fixing
the local algorithm might have annihilated thermal Dirac loops of different
orientation to a large extent. This is expressed by the ratio~\eq{eq:r3}
which is of the order of 80 \% or larger. This would correspond to a
suppression of so called double Dirac loops (analogues of the mentioned
DDS in four dimensions) in the configurations possessing two oppositely
oriented TDL's.

\begin{figure}[!htb]
\centerline{\includegraphics[angle=0,scale=0.5,clip=true]{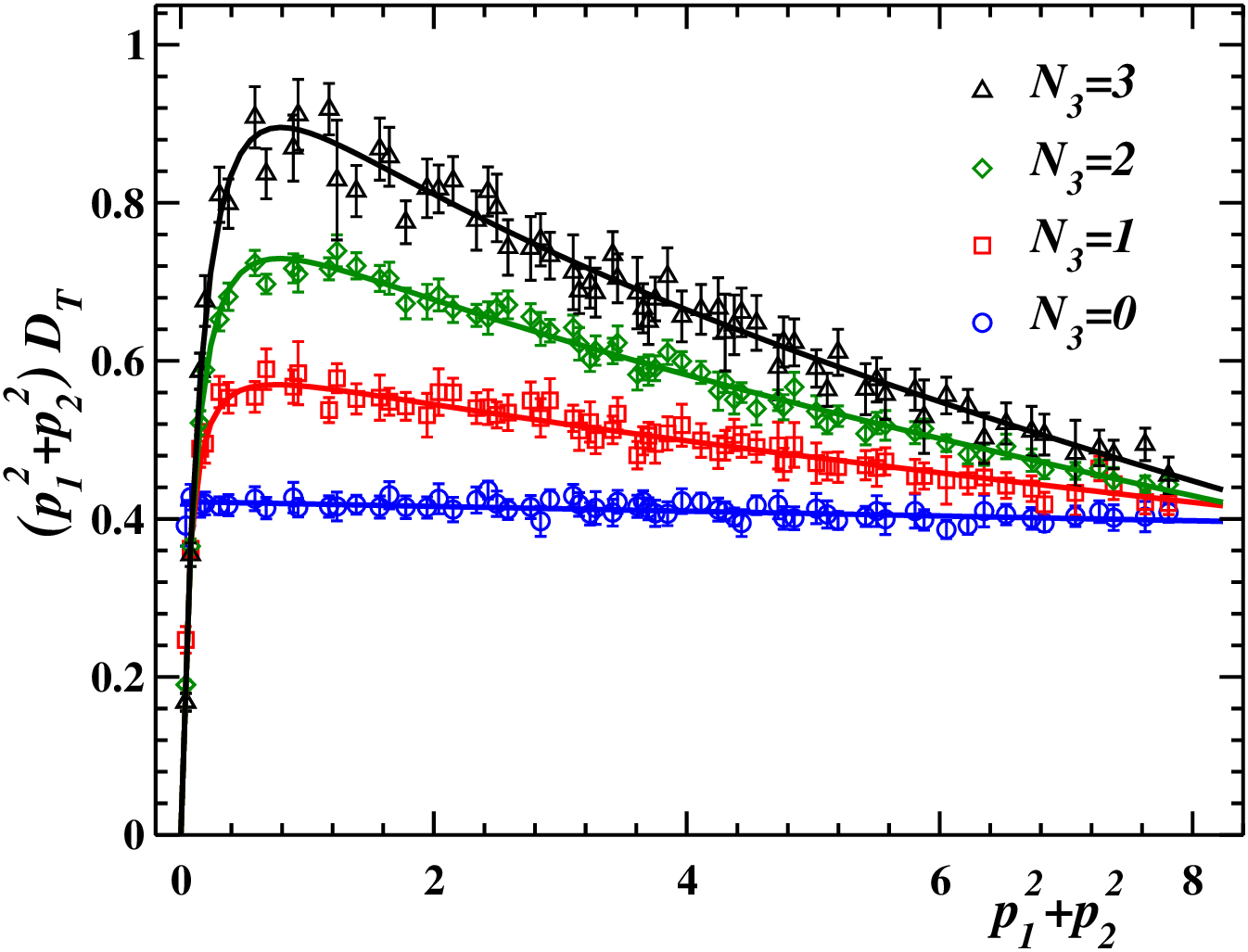}}
  \caption{The total formfactor $D_T$ for spatial photons in the different TDL      sectors, using the blended update.
   The data are fitted by the function~\eq{def:anomalous_fit}.}
  \label{fig:total:fits}
\end{figure}

It turns our that the fit of the formfactor $D_T$
using the functional dependence on $p$ given in~\eq{def:anomalous_fit}
works very well separately for all sectors
$N_3 = 0, \dots, 4$,
with $\chi^2/d.o.f. \approx 0.5$.
The corresponding fitting curves yielding $\alpha_T$ and $m_T$
separately for each
$N_3=0,\dots,3$, are shown in Figure~\ref{fig:total:fits}.
%

\begin{figure}[!htb]
  \begin{center}
    \begin{tabular}{cc}
    \includegraphics[angle=0,scale=0.35,clip=true]{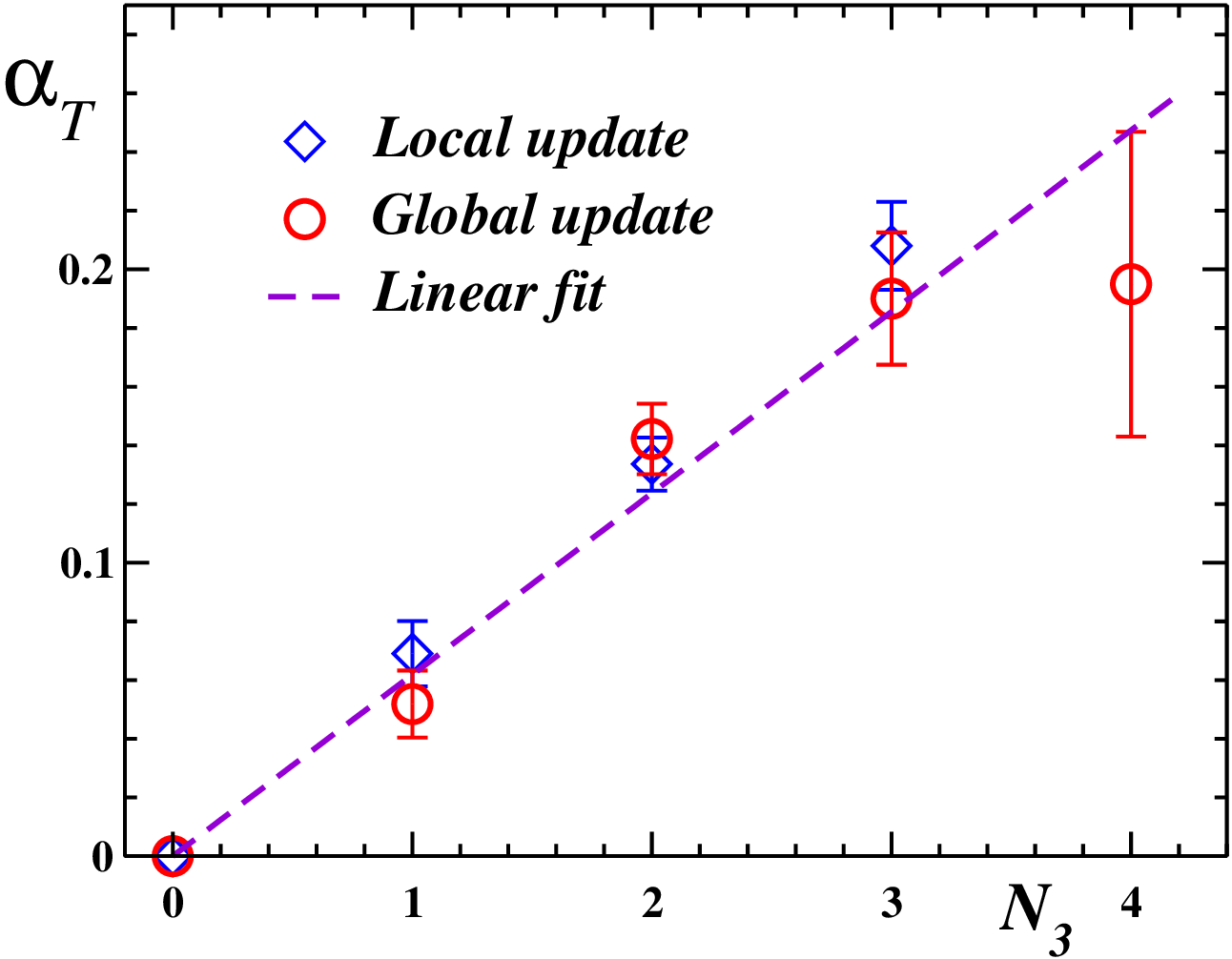} &
      \hspace{5mm}
    \includegraphics[angle=0,scale=0.35,clip=true]{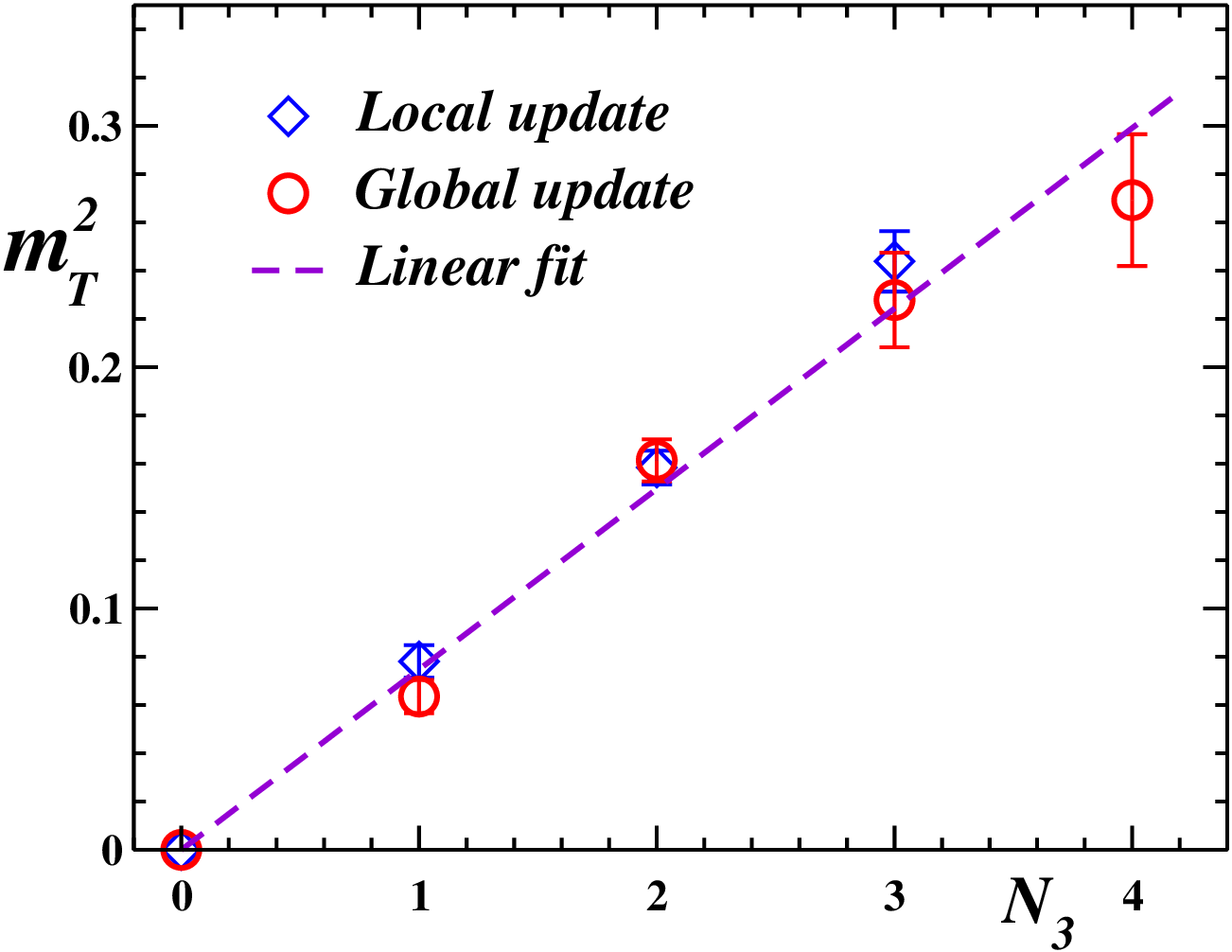} \\
      (a) & \hspace{5mm}  (b)
\vspace{5mm}
      \\
    \includegraphics[angle=0,scale=0.35,clip=true]{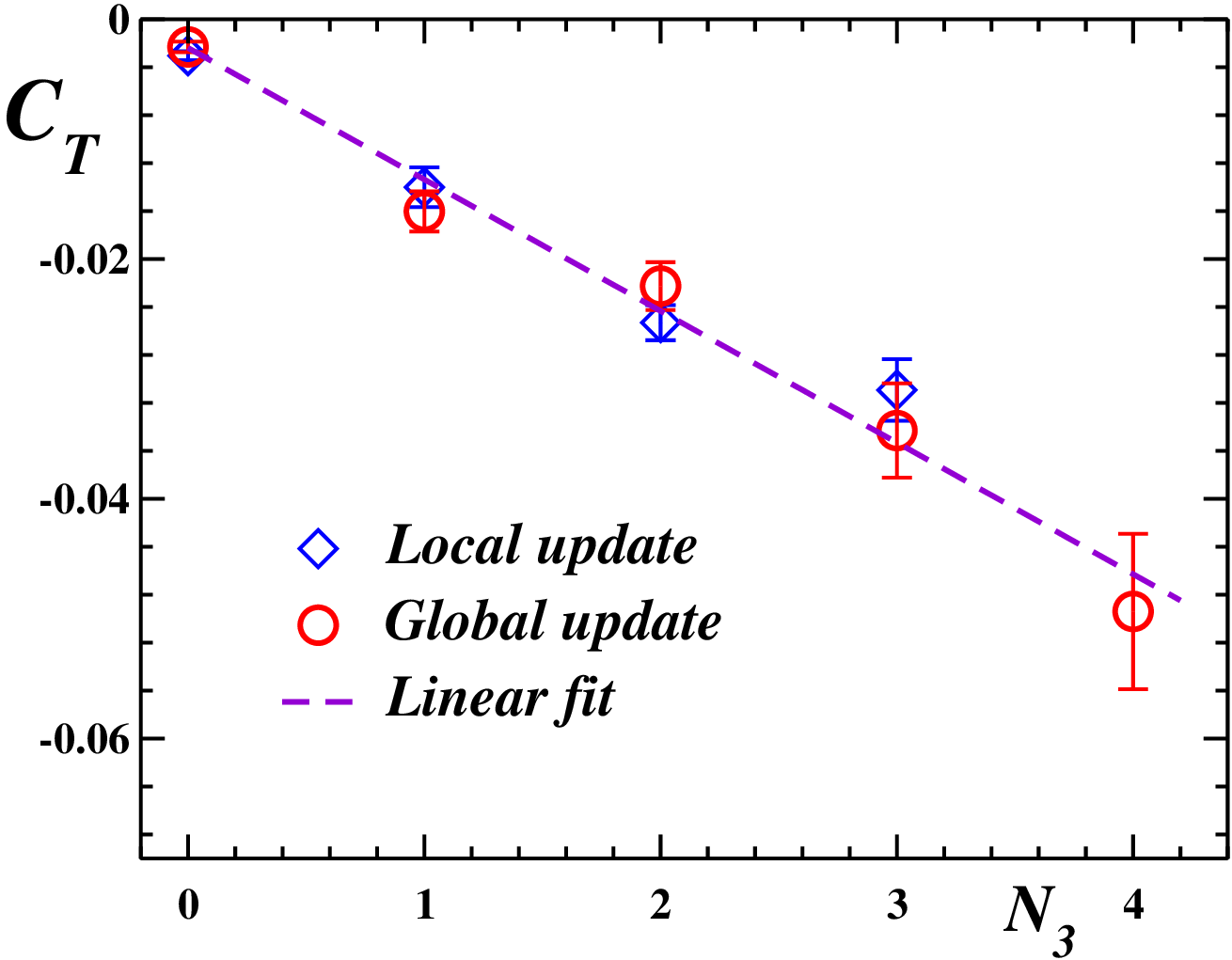} &
      \hspace{5mm}
    \includegraphics[angle=0,scale=0.35,clip=true]{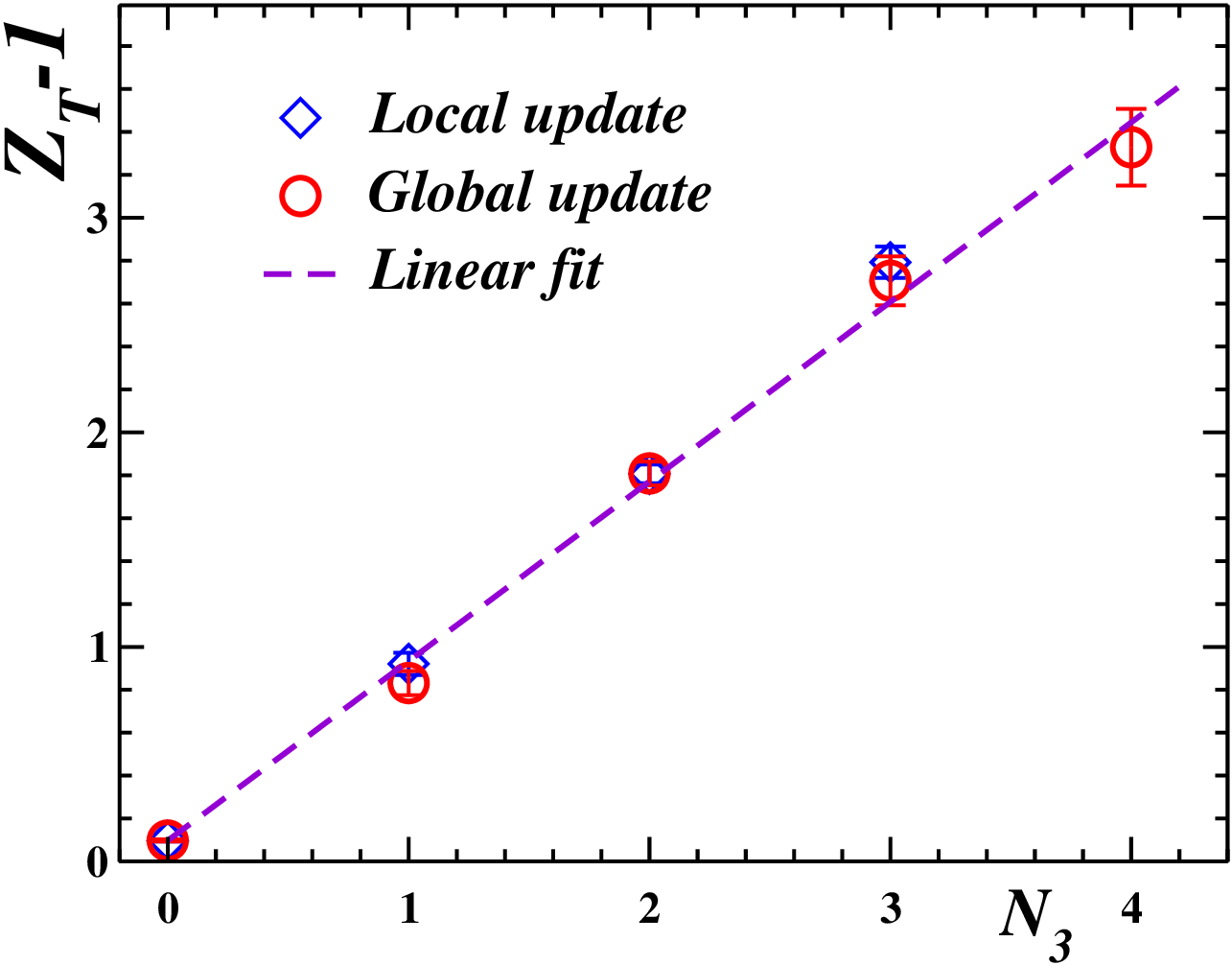} \\
      (c) & \hspace{5mm}  (d)
    \end{tabular}
  \end{center}
  \caption{Fit parameters $\alpha_T$, $m_T$, $Z_T$ and $C_T$
  for the $D_T$ formfactor {\it vs.} the number of thermal Dirac
  loops $N_3$ using data obtained with blended or local updates.
  The fits by Eqs.~\eq{eq:fit.of.fit.parameters} are presented by
  dashed lines.}
  \label{fig:fit:parameters}
\end{figure}

The fitting parameters themselves as functions of the TDL
multiplicity $N_3$ are shown in
Figure~\ref{fig:fit:parameters} both for the local and the
blended updates.
%
First we notice that the fit parameters in a given loop sector are
practically independent of the update.
As one can see from Table~\ref{tbl:statistics}, our statistics for $N_3=4$
achieved by exclusively local updates is quite low such that we omit the
corresponding points in Figures~\ref{fig:fit:parameters}. We emphasize the
observation that
it is extremely unlikely to produce, by updates without global changes of flux,
configurations which finally, after gauge fixing, end up in the sector $N_3=4$.

In the sector without thermal Dirac loops ($N_3=0$)
the anomalous dimension $\alpha_T$, the mass parameter $m_T$ and the contact
term parameter $C_T$ are consistent with zero.
At the same time, the renormalization of the photon wavefunction $Z_T$ is very
close to unity. Thus, if we would restrict ourselves to the $N_3=0$ sector in
the propagator measurements,
there is practically no difference between the $D_L$ and $D_T$ formfactors
in the deconfinement phase, and we
reproduce the expected behavior of a trivial free photon propagator.

In the non-zero TDL sector ($N_3 \geqslant 1$) the parameters
$\alpha_T$, $m_T$ and $C_T$ become non-zero, and the parameter $Z_T$ deviates
from unity. Thus, the formfactor acquires a non-trivial momentum dependence
compared to the sector without thermal Dirac loops.
The dependence of the propagator parameters, $\alpha_T$, $m_T^2$,
$Z_T$ and $C_T$, on the multiplicity of thermal Dirac loops,
$N_3$, can be described by simple linear functions:
\beqn
  \alpha_T = a_\alpha \, N_3\,,\quad m_T^2 = a_m N_3\,,\quad
   Z_T - 1 =a_Z N_3\,, \quad C_T = a_C N_3\,.
  \label{eq:fit.of.fit.parameters}
\eeqn
The corresponding fits are shown by the dashed lines in
Figures~\ref{fig:fit:parameters}.
The quality of these fits is approximately the same,
$\chi^2/d.o.f. \approx 1.5$.
The proportionality coefficients are $a_\alpha=0.062(5)$,
$a_m=0.0745(38)$, $a_Z=0.84(3)$ and $a_C=-0.011(1)$.
We recall that all this refers to a temperature corresponding to $\beta=2.6$.
We expect that in general these coefficients must depend on temperature.

\section{Conclusions}

We have studied the influence of thermal Dirac loops
on the gauge boson propagator within the finite temperature compact $U(1)$
gauge model in three dimensions. We have used the Landau gauge and
worked deep inside the deconfinement phase where monopoles are very dilute
and Dirac strings are dilute as well.
This allows us to unambiguously recognize the {\it thermal Dirac loops}.
Wrapping Dirac strings are ubiquitous on a finite lattice.
On an asymmetric lattice at high temperature they are predominantly closed
along the temperature direction which is the shortest lattice direction.
Although closed Dirac loops along the spatial directions ({\it non-thermal
Dirac loops}) are not completely excluded, they are extremely rare and do
not yield an essential contribution to the propagator $D_T$ in question.

Strictly speaking, in the Landau gauge
Dirac strings closed due to periodic boundary conditions are
artifacts of the gauge fixing, because the Landau gauge condition corresponds
to the minimization of the number and length of Dirac
strings~\cite{Chernodub:2002gp}.
We have found that the presence of such thermal Dirac loops
seriously affects the properties of the propagator, in the considered
case those of $D_T$, in the deconfinement phase.
The propagator formfactor corresponding to spatial photons in a sector
with a non-vanishing number $N_3$ of TDL's mimics a momentum dependence similar to what is known from the confinement phase, Eq.~\eq{def:anomalous_fit}.
The parameters, which describe the deviation from the free photon propagator,
which ought to be expected in the deconfinement phase, are found clearly
proportional to $N_3$ because we were working in dilute gas regime.
An explanation of this coincidence may lie in the fact that the monopoles,
which are active in the confinement phase, contribute to the propagator only
indirectly, {\it i.e.} via the Dirac strings.

\begin{acknowledgments}
M.~N.~Ch. acknowledges a partial support from the grants
RFBR 01-02-17456, DFG 436 RUS 113/73910 and RFBR-DFG 03-02-04016.
E.-M.~I. is supported by DFG through the DFG-Forschergruppe
''Lattice Hadron Phenomenology'' (FOR 465). M.~N.~Ch. and E.-M.~I. feel much
obliged for a kind hospitality extended to them by the theory group of the
Kanazawa University where this work was completed.
\end{acknowledgments}

\end{document}